\documentclass[aps,twocolumn,amsmath,amssymb,superscriptaddress,prb]{revtex4}

\usepackage{graphicx}
\usepackage{bm}
\usepackage[dvips]{color}

\begin{document}

\title{Observation of the inverse spin Hall effect in silicon}

\author{Kazuya Ando\footnote{ando@imr.tohoku.ac.jp}}
\affiliation{Institute for Materials Research, Tohoku University, Sendai 980-8577, Japan}

\author{Eiji Saitoh}
\affiliation{Institute for Materials Research, Tohoku University, Sendai 980-8577, Japan}
\affiliation{CREST, Japan Science and Technology Agency, Sanbancho, Tokyo 102-0075, Japan}
\affiliation{The Advanced Science Research Center, Japan Atomic Energy Agency, Tokai 319-1195, Japan }

\maketitle

\textbf{
The spin-orbit interaction in a solid couples the spin of an electron to its momentum. This coupling gives rise to mutual conversion between spin and charge currents: the direct and inverse spin Hall effects. The spin Hall effects have been observed in metals and semiconductors. However, the spin/charge conversion has not been realized in one of the most fundamental semiconductors, silicon, where accessing the spin Hall effects has been believed to be difficult because of its very weak spin-orbit interaction. Here we report observation of the inverse spin Hall effect in silicon at room temperature. The spin/charge current conversion efficiency, the spin Hall angle, is obtained as 0.0001 for a p-type silicon film. In spite of the small spin Hall angle, we found a clear electric voltage due to the inverse spin Hall effect in the p-Si film, demonstrating that silicon can be used as a spin-current detector. 
}

Silicon is a group I\hspace{-.1em}V semiconductor having the diamond structure. This material has played a crucial role in exploring the physics of semiconductors. Silicon has been broadly viewed as an ideal host also for spintronics due to its low atomic mass, crystal inversion symmetry, and near lack of nuclear spin, resulting in the exceptionally long spin lifetime~\cite{Zutic,ActaSS,PhysRevLett.104.016601}. 

Along with long spin lifetimes, key elements for spintronics are generation and detection of spin currents~\cite{Wolf,PhysRevLett.55.1790,Jedema,citeulike:1629854,S.A.Crooker09302005,PhysRevLett.96.176603,PhysRevLett.102.036601,PhysRevLett.87.016601}. A promising method is the utilization of the direct and inverse spin Hall effects (DSHE/ISHE), which convert a charge current into a spin current and vice versa~\cite{Bakun,Dyakonov,Hirsch,Murakami,Sinova,Kato,Wunderlich,Sih2,Stern,TimeSHE,Wunderlich24122010,PhysRevLett.96.056601,PhysRevLett.95.166605,Saitoh,KimuraPRL,Valenzuela,Matsuzaka,PhysRevLett.105.156602,Ballistic}. However, the underlying origin of the spin Hall effects is the spin-orbit interaction and thus it is natural to expect that the spin Hall effects are not accessible in a material which shows long spin lifetimes, such as silicon. Generation of spin currents from an electric field via the spin-orbit interaction, the DSHE, was first observed in GaAs using optical detection techniques, a Kerr-microscopy and a two-dimensional light-emitting diode~\cite{Kato,Wunderlich}. Although these optical techniques have played a crucial role for investigating the physics of the DSHE~\cite{Kato,Wunderlich,Sih2,TimeSHE,Matsuzaka}, the application range of the techniques is limited to direct bandgap semiconductors with strong spin-orbit interaction; the indirect bandgap of silicon precludes using these techniques, making it difficult to explore the DSHE in silicon along with its very weak spin-orbit interaction. 

\begin{figure}[bt]
\includegraphics[scale=1]{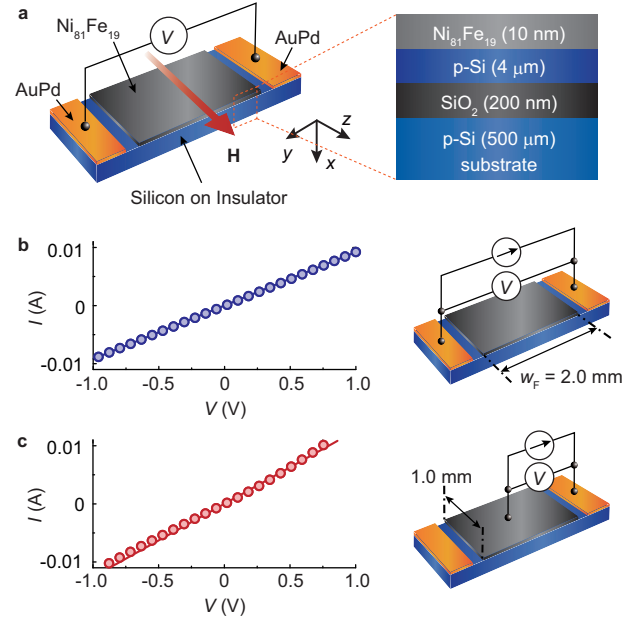}
\caption{{\bfseries Experimental setup.} \textbf{a}, A schematic illustration of the Ni$_{81}$Fe$_{19}$/p-Si film used in this study. ${\bf H}$ represents an external magnetic field. \textbf{b}, Current-voltage ($I$-$V$) characteristic measured for the Ni$_{81}$Fe$_{19}$/p-Si film, where the two electrodes are attached to the AuPd layers. $w_\text{F}$ is the length of the Ni$_{81}$Fe$_{19}$ layer. \textbf{c}, $I$-$V$ characteristic measured for the Ni$_{81}$Fe$_{19}$/p-Si film. The two electrodes are attached to the Ni$_{81}$Fe$_{19}$ layer and AuPd layer, respectively.  }
\label{fig1} 
\end{figure}

\begin{figure}[tb]
\includegraphics[scale=1]{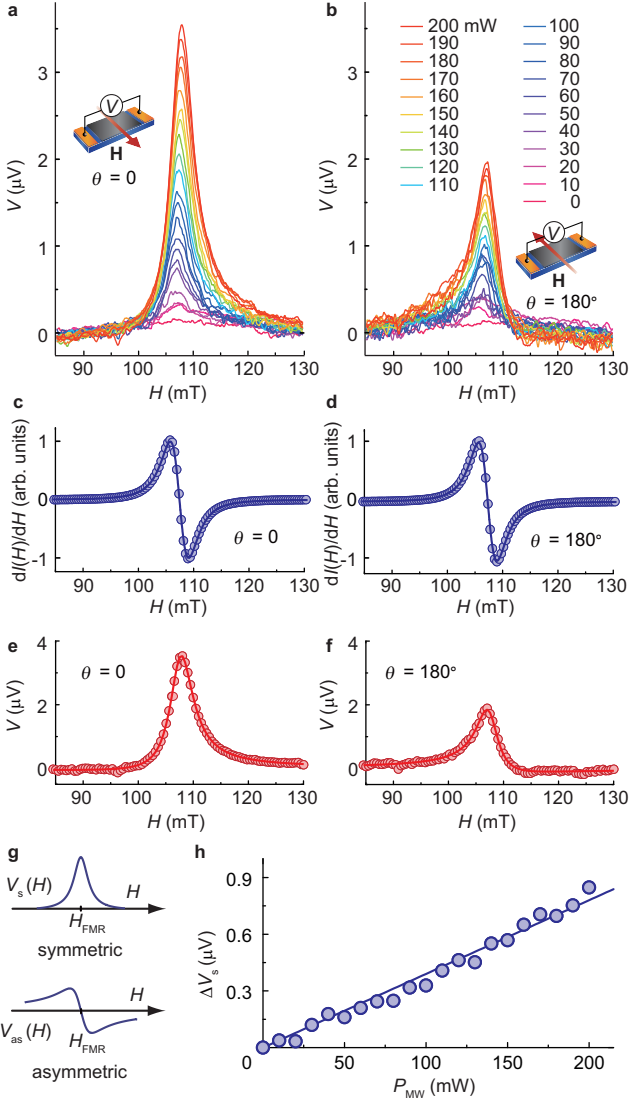}
\caption{{\bfseries Observation of ISHE in silicon.} \textbf{a}, Field ($H$) dependence of the electromotive force $V$ measured for the Ni$_{81}$Fe$_{19}$/p-Si film when $\theta=0$ at different microwave excitation powers. The inset shows a schematic illustration of the experimental setup when $\theta=0$. \textbf{b}, $H$ dependence of $V$ measured when $\theta=180^{\circ}$. \textbf{c}, $H$ dependence of the FMR signal $dI(H)/dH$ when $\theta=0$ at 200 mW. $I$ is the microwave absorption intensity. %The solid circles are the experimental data. The solid curve shows the fitting result using the first derivative of a Lorentz function. 
\textbf{d}, $H$ dependence of $dI(H)/dH$ when $\theta=180^\circ$ at 200 mW. \textbf{e}, $H$ dependence of $V$ when $\theta=0$. The solid circles are the experimental data. The solid curve shows the fitting result. \textbf{f}, $H$ dependence of $V$ when $\theta=180^{\circ}$. \textbf{g}, The spectral shape of the symmetric $V_\text{s}(H)$ and asymmetric $V_\text{as}(H)$ components of the electromotive force $V(H)$. \textbf{h}, Microwave power $P_\text{MW}$ dependence of $\Delta V_\text{s}$. 
}
\label{fig2} 
\end{figure}

The spin-orbit interaction responsible for the DSHE also causes the conversion of a spin current into an electric field, the ISHE~\cite{Saitoh,KimuraPRL,Valenzuela}, which could offer a way to circumvent the above obstacle in exploring the spin Hall effects. The ISHE enables the electric measurement of the spin/charge conversion through the spin-orbit interaction, as demonstrated, for example, in platinum and gallium arsenide~\cite{Saitoh,KimuraPRL,Wunderlich24122010,PhysRevLett.105.156602}. The electric field ${\bf E}_\text{ISHE}$ generated by the ISHE from a spin current ${\bf j}_\text{s}$ with the spin-polarization vector $\bm{\sigma}$ is described as~\cite{Saitoh}
\begin{equation}
{\bf E}_{\rm ISHE} =\left(\theta_\text{SHE} \rho_\text{N} \right) {\bf j}_\text{s}\times \bm{\sigma},\label{ISHEeq}
\end{equation}
where $\theta_\text{SHE}=\sigma_\text{SHE}/\sigma_\text{N}$ is the spin Hall angle, $\sigma_\text{SHE}$ and $\sigma_\text{N}$ are the spin Hall conductivity and electric conductivity, respectively, and $\rho_\text{N}$ is the electric resistivity. Equation~(\ref{ISHEeq}) shows that the magnitude of the electric field due to the ISHE is proportional to the resistivity $\rho_\text{N}$ of the material, indicating that the ISHE enables sensitive detection of spin currents especially in high resistivity materials, such as semiconductors.

Although spin injection efficiency into semiconductors is drastically limited by the impedance mismatch problem~\cite{Schmidt}, recent advances revealed that efficient spin injection is possible using hot-electron injection~\cite{Appelbaum}, tunnel barriers~\cite{Jonker,Dash}, and spin pumping~\cite{AndoNMad}. In particular, the generation of spin currents from magnetization precession~\cite{Tserkovnyak1,Brataas}, a recently discovered method utilizing spin pumping, enables high-density spin current injection into a macroscopic area~\cite{AndoNMad}, which is difficult to achieve by other methods. This is beneficial for enhancing the electric voltage $V_\text{ISHE}=w_\text{F}E_\text{ISHE}$ due to the ISHE; $V_\text{ISHE}$ is proportional both to the spin current density $j_\text{s}$ and length $w_\text{F}$ of the sample along ${\bf E}_\text{ISHE}$. The combination of the spin pumping and ISHE has been observed and is a well established technique in metallic systems~\cite{AndoJAPfull}. This has also been applied to semiconductors with strong spin-orbit interaction, enabling the observation of the ISHE in heavily doped n- and p-type GaAs~\cite{AndoNMad}. In this work, we experimentally demonstrate that the combination of the ISHE and spin pumping provides a route for exploring the spin/charge current conversion in high resistivity materials with weak spin-orbit interaction by showing successful measurement of the ISHE in silicon at room temperature.

\begin{figure*}[tb]
\includegraphics[scale=1]{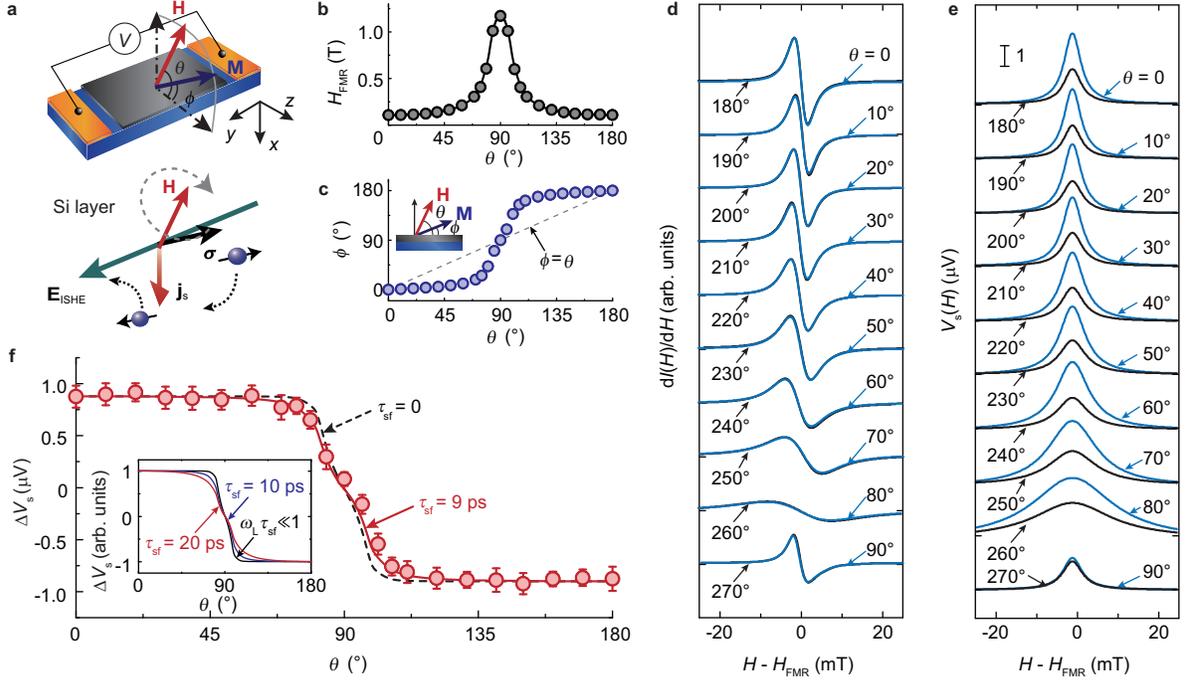}
\caption{{\bfseries Angular dependence of ISHE signal and spin precession.} \textbf{a}, A schematic illustration of the Ni$_{81}$Fe$_{19}$/p-Si film when the external magnetic field ${\bf H}$ is applied oblique to the film plane. ${\bf M}$ denotes the static component of the magnetization. $\theta$ and $\phi$ show the magnetic field angle and magnetization angle, respectively. \textbf{b}, Magnetic field angle $\theta$ dependence of the ferromagnetic resonance field $H_{\rm FMR}$ measured for the Ni$_{81}$Fe$_{19}$/p-Si film. The filled circles represent the experimental data. The solid curve is the numerical solution of the Landau-Lifshitz-Gilbert equation with the saturation magnetization $4\pi M_s=0.852$ T. \textbf{c}, Magnetic field angle $\theta$ dependence of the magnetization angle $\phi$ for the Ni$_{81}$Fe$_{19}$/p-Si film estimated using the Landau-Lifshitz-Gilbert equation with the measured values of $H_\text{FMR}$. \textbf{d}, Magnetic-field-angle $\theta$ dependence of the FMR signal $dI(H)/dH$ measured for the Ni$_{81}$Fe$_{19}$/p-Si film at 200 mW. \textbf{e}, Magnetic-field-angle $\theta$ dependence of the symmetric component of the electromotive force $V_\text{s}(H)$ extracted by a fitting procedure from the measured $V$ spectra for the Ni$_{81}$Fe$_{19}$/p-Si film at 200 mW. \textbf{f}, Magnetic field angle $\theta$ dependence of $\Delta V_\text{s}$. The solid circles are the experimental data. The solid curve is the theoretical curve obtained from equation~(\ref{hanle}) with $\tau_\text{sf}=9$ ps. The dashed curve is the theoretical curve for $\tau_\text{sf}=0$. The error bars represent the 90\% confidence interval. The inset shows the $\theta$ dependence of $\Delta V_\text{s}$ calculated from equation~(\ref{hanle}) with $\tau_\text{sf}=20$ ps (the red curve), $\tau_\text{sf}=10$ ps (the blue curve), and $\omega_\text{L}\tau_\text{sf}\ll1$ (the black curve). }
\label{fig3} 
\end{figure*}

\begin{figure}[tb]
\includegraphics[scale=1]{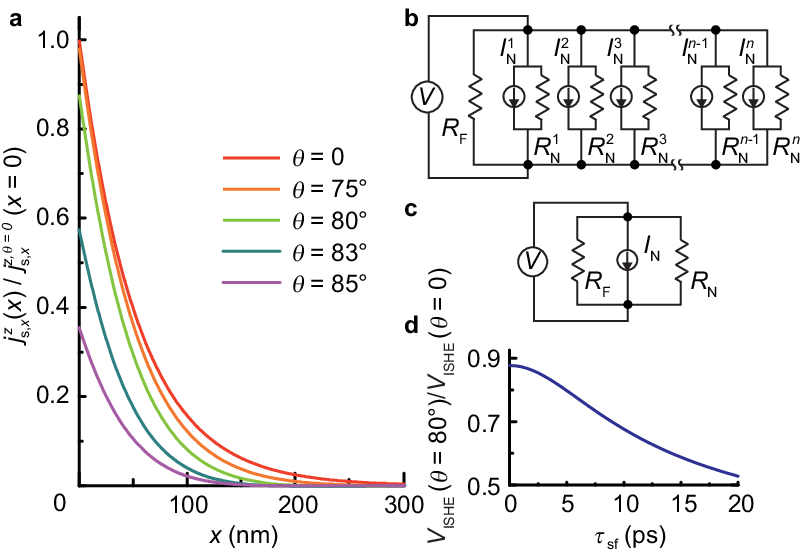}
\caption{{\bfseries Spin current relaxation.} \textbf{a}, The spin current density $j_{\text{s},x}^z (x)$ generated by the spin pumping for $\tau_\text{sf}=9$ ps. Here, $j_{\text{s},x}^{z,\theta=0} (x=0)$ is the spin current density at the interface when the external magnetic field is applied along the film plane ($\theta=0$). The parameters used for the calculation are shown in the text. \textbf{b}, An equivalent circuit model of the Ni$_{81}$Fe$_{19}$/p-Si film. $R_\text{F}$ is the electrical resistance of the Ni$_{81}$Fe$_{19}$ layer. \textbf{c}, A simplified equivalent circuit model of the Ni$_{81}$Fe$_{19}$/p-Si film. \textbf{d}, The spin relaxation time $\tau_\text{sf}$ dependence of the ISHE signal $V_\text{ISHE}$ at $\theta=80^\circ$ calculated from equation~(\ref{hanle}). }
\label{fig4} 
\end{figure}

\bigskip\noindent
\textbf{Results}\\
\textbf{Detection of inverse spin Hall effect in silicon.} Figure~\ref{fig1}a shows a schematic illustration of the sample used in this study. The sample is a Ni$_{81}$Fe$_{19}$/B-doped Si (Ni$_{81}$Fe$_{19}$/p-Si) film with a doping concentration of $N_\text{A}=2\times 10^{19}$ $\text{cm}^{-3}$ (for details, see Methods). Two ohmic contacts were attached on the p-Si layer (see Figs.~\ref{fig1}a and \ref{fig1}b). Here, note that the current-voltage characteristic shown in Fig.~\ref{fig1}c shows an almost ohmic behavior at the Ni$_{81}$Fe$_{19}$/p-Si interface, suggesting strong dynamical exchange interaction $J_\text{ex}$ between the magnetization in the Ni$_{81}$Fe$_{19}$ layer and carrier spins in the p-Si layer~\cite{AndoNMad}.

We measured the ferromagnetic resonance (FMR) signal and electric-potential difference $V$ between the electrodes attached to the p-Si layer to detect the ISHE~\cite{AndoNMad}; in the FMR condition, the spin pumping driven by the dynamical exchange interaction injects pure spin currents into the p-Si layer. This spin current gives rise to an electric voltage $V_{\rm ISHE}$ via the ISHE in the p-Si layer. During the measurements, the Ni$_{81}$Fe$_{19}$/p-Si sample was placed at the center of a TE$_{011}$ microwave cavity with the frequency of $f=9.45$ GHz, where the microwave magnetic field was applied along the $y$ direction (see Fig.~\ref{fig1}a). An external static magnetic field ${\bf H}$ was applied along the film plane as shown in Fig.~\ref{fig1}a. All of the measurements were performed at room temperature.

Figures~\ref{fig2}a and \ref{fig2}b show the d.c. electromotive force signals measured for the Ni$_{81}$Fe$_{19}$/p-Si film at various microwave excitation power when the external magnetic field $\bf{H}$ is applied along the film plane at $\theta=0$ and $\theta=180^\circ$ (see the insets), respectively. Here, $\theta$ is the out-of-plane angle of ${\bf H}$. In the $V$ spectra, clear electromotive force signals are observed around the ferromagnetic resonance field $H_\text{FMR}$ (compare the $V$ spectra with the FMR spectra shown in Figs.~\ref{fig2}c and 2d). Figures~\ref{fig2}c and \ref{fig2}d show that the microwave absorption intensity $I$ is identical for $\theta=0$ and $180^{\circ}$. In contrast, importantly, the magnitude of the electromotive force $V$ is clearly changed by reversing the magnetic field direction as shown in Figs.~\ref{fig2}a and \ref{fig2}b; this distinctive behavior of $V$ is the key feature of the ISHE induced by the spin pumping~\cite{AndoNMad}.

The electromotive force observed here is the combination of the ISHE in the p-Si layer, the ordinary Hall effect (OHE) in the p-Si layer, the anomalous Hall effect (AHE) in the Ni$_{81}$Fe$_{19}$ layer, and heating effects. The direct contribution from the ISHE in the p-Si layer can be extracted as follows (see Methods). The OHE and AHE voltages can be ruled out from the observed electromotive force by fitting the $V$ spectra using a combination of symmetric $V_\text{s}(H)=V_\text{s}{\Gamma^2}/\left[{(H-H_\text{FMR})^2+\Gamma^2}\right]$ (absorption shape) and asymmetric $V_\text{as}(H)=V_\text{as}\left[-2\Gamma(H-H_\text{FMR})\right]/\left[{(H-H_\text{FMR})^2+\Gamma^2}\right]$ (dispersion shape) functions~\cite{Saitoh}, $V(H)=V_\text{s}(H)+V_\text{as}(H)$, where $H_\text{FMR}$ is the resonance field. Figures~\ref{fig2}e and \ref{fig2}f are the fitting result for the $V$ spectra at 200 mW for $\theta=0$ and $\theta=180^\circ$, respectively, showing that the observed $V$ spectra are well reproduced with $V_\text{s}= 3.50$ $\mu$V and $V_\text{as}=-0.41$ $\mu$V for $\theta=0$ and $V_\text{s}= 1.76$ $\mu$V and $V_\text{as}=0.41$ $\mu$V for $\theta=180^\circ$. What is notable is that the Hall voltage due to rectification changes its sign across $H_\text{FMR}$ as shown in Fig.~\ref{fig2}g (Ref.~\onlinecite{Saitoh}). In contrast, the electromotive force due to the ISHE is proportional to the microwave absorption intensity~\cite{AndoJAPfull}. Here, $V_\text{s}$ is attributed both to the ISHE in the p-Si layer and heating effects~\cite{AndoNMad}. To eliminate the heating effects arising from the microwave absorption from the $V$ spectra, we define $\Delta V_\text{s}=\left( V_\text{s}(\theta)-V_\text{s}(\theta+180^\circ)\right)/2$, since the ISHE voltage due to the spin pumping changes its sign by reversing $\bf{H}$ while the electromotive force due to the heating effects is independent on the $\bf{H}$ direction. In Fig.~\ref{fig2}h, we show the microwave power $P_\text{MW}$ dependence of $\Delta V_\text{s}$. $\Delta V_\text{s}$ increases linearly with $P_\text{MW}$, as expected for the ISHE induced by the spin pumping~\cite{AndoJAPfull}. $\Delta V_\text{s}$ signal disappears when an in-plane magnetic field is applied parallel to the direction across the electrodes, supporting that $\Delta V_\text{s}$ is attributed to the ISHE in the p-Si layer because of equation~(\ref{ISHEeq}).

\bigskip\noindent
\textbf{Spin precession and inverse spin Hall effect.} To further buttress the above result, we measured the out-of-plane magnetic field angle $\theta$ dependence of $\Delta V_\text{s}$, which provides further evidence that the observed $\Delta V_\text{s}$ signals are attributed to the ISHE induced by the spin injection in the p-Si layer. Here, the out-of-plane magnetic field angle $\theta$ is defined in Fig.~\ref{fig3}a. As shown in Fig.~\ref{fig3}a, when ${\bf H}$ is applied oblique to the film plane, the magnetization precession axis is not parallel to ${\bf H}$ due to the demagnetization field in the Ni$_{81}$Fe$_{19}$ layer. The relation between the external magnetic field angle $\theta$ and the angle of magnetization-precession axis $\phi$ can be obtained using the Landau-Lifshitz-Gilbert equation with the measured values of the resonance field $H_\text{FMR}$ shown in Fig.~\ref{fig3}b (Ref.~\onlinecite{AndoJAPfull}). The $\theta$ dependence of $\phi$ for the Ni$_{81}$Fe$_{19}$/p-Si film is shown in Fig.~\ref{fig3}c. In Figs.~\ref{fig3}d and \ref{fig3}e, we show the $\text{d}I/\text{d}H$ and $V_\text{s}(H)$ signals for the Ni$_{81}$Fe$_{19}$/p-Si film at different $\theta$. As shown in Fig.~\ref{fig3}e, $\Delta V_\text{s}$ disappears when the external magnetic field is applied perpendicular to the film plane; the $\theta$ dependence of $\Delta V_\text{s}$ shows the drastic variation of $\Delta V_\text{s}$ around $\theta=90^\circ$ (see in Fig.~\ref{fig3}f ). Here, note that the spin-polarization vector $\bm{\sigma}$ of the spin current injected into the p-Si layer is parallel to the magnetization-precession axis. Therefore, the spins of the spin current precess around the axis parallel to ${\bf H}$ as shown in Fig.~\ref{fig3}a. This is described in the Bloch equation with spin diffusion and precession in the p-Si layer:
\begin{widetext}
\begin{equation}
\frac{{\partial {\bf{m}}(x,t)}}{{\partial t}} =  - \gamma _{\rm{c}} \left[ {{\bf{m}}(x,t) \times {\bf{H}}} \right] - \frac{{{\bf{m}}(x,t)}}{{\tau _{{\rm{sf}}} }} + D_{\rm{N}} \nabla ^2 {\bf{m}}(x,t) + 2\left( {j_{{\text{s}},x}^x  {\bf{e}}_x  + j_{{\text{s}},x}^z {\bf{e}}_z } \right)\delta (x), \label{Bloch}
\end{equation}
\end{widetext}
where ${\bf{m}}(x,t)$ is the magnetization of carriers in the p-Si layer; $\gamma_\text{c}$ and $\tau_\text{sf}$ are the gyromagnetic ratio and spin relaxation time of carriers in the p-Si layer, respectively; $D_\text{N}$ is the diffusion constant in the p-Si layer; ${\bf e}_x$ and ${\bf e}_z$ are the unit vector parallel to the $x$ and $z$ axes, respectively (see Fig.~\ref{fig3}a); $\delta (x)$ is the delta function and $j_{\text{s},q}^p$ is the spin current density with the spin orientation direction $p$ and flow direction $q$ at the interface $x=0$. Thus $j_{{\text{s}},x}^x=-j_\text{s}\sin\phi$ and $j_{{\text{s}},x}^z=j_\text{s}\cos\phi$, where the spin current density $j_\text{s}$ generated by the spin pumping at the interface in the FMR condition is given by~\cite{AndoJAPfull},
\begin{widetext}
\begin{equation}
j_\text{s}=\frac{ g_\text{r}^{\uparrow\downarrow}  \gamma^2 h^2 \hbar \left[4 \pi  M_\text{s} \gamma  \cos ^2\phi+\sqrt{ (4 \pi M_\text{s})^2 \gamma ^2 \cos ^4\phi+4 \omega ^2}\right]}{8 \pi \alpha^2  \left((4 \pi  M_\text{s})^2 \gamma ^2 \cos ^4\phi +4 \omega ^2\right)}.\label{pumping2}
\end{equation}
\end{widetext}
Here, $g_\text{r}^{\uparrow\downarrow}$ is the spin mixing conductance, $\gamma$ and $M_\text{s}$ are the gyromagnetic ratio and saturation value of the magnetization ${\bf M}$, respectively, $\alpha$ is the Gilbert damping constant, $h$ is the microwave magnetic field, and $\omega = 2\pi f$ is the angular frequency of the magnetization precession. By solving equation~(\ref{Bloch}) for the equilibrium condition ($\partial {\bf m}/\partial t =0$), we obtain 
\begin{widetext}
\begin{equation}
\left( {\begin{array}{*{20}{c}}
{j_{\text{s},x}^x(x)}\\
{j_{\text{s},x}^y(x)}\\
{j_{\text{s},x}^z(x)}
\end{array}} \right) = \left( {\begin{array}{*{20}{c}}
{{-j_\text{s}}\sin \theta \cos (\theta  - \phi ){e^{ - x/{\lambda _{\rm{N}}}}} + {j_\text{s}}\cos \theta \sin (\theta  - \phi ){\rm{Re}}\left[ {{e^{ - x/{\lambda _\omega }}}} \right]}\\
{{-j_\text{s}}\sin (\theta  - \phi ){\rm{Im}}\left[ {{e^{ - x/{\lambda _\omega }}}} \right]}\\
{{j_\text{s}}\cos \theta \cos (\theta  - \phi ){e^{ - x/{\lambda _{\rm{N}}}}} + {j_\text{s}}\sin \theta \sin (\theta  - \phi ){\rm{Re}}\left[ {{e^{ - x/{\lambda _\omega }}}} \right]}
\end{array}} \right),
\end{equation}
\end{widetext}
where $\lambda_\omega=\lambda_\text{N}/\sqrt{1+i \omega_\text{L}\tau_\text{sf}}$ and $\omega_\text{L}=\gamma_\text{c} H_\text{FMR}$; $\lambda_\text{N}=\sqrt{D_\text{N}\tau_\text{sf}}$ is the spin diffusion length of the Si layer; ${\mathop{\rm Re}\nolimits} \left[ {{e^{ - x/{\lambda _\omega }}}}\right]$ and ${\mathop{\rm Im}\nolimits} \left[ {{e^{ - x/{\lambda _\omega }}}} \right]$ are the real and imaginary part of $e^{ - x/{\lambda _\omega }}$, respectively. Since the spin current flows along the $x$ direction, the electric field induced by the ISHE, $E_\text{ISHE}(x)$, is proportional to $j_{\text{s},x}^z(x)$. As shown in Fig.~\ref{fig4}a, $j_{\text{s},x}^z (x)$ decays because of the spin relaxation in the p-Si layer; the charge current density $j_\text{c}(x)$ generated by the ISHE also depends on $x$, which induces short circuit currents in the Ni$_{81}$Fe$_{19}$ and p-Si layers~\cite{AndoJAPfull}. Equivalent circuit models of the Ni$_{81}$Fe$_{19}$/p-Si film is shown in Figs.~\ref{fig4}b and \ref{fig4}c (see Methods). Therefore, we obtain the angular dependence of the ISHE signal ${\rm{   }}V_{{\rm{ISHE}}}$ in the presence of spin precession as~\cite{AndoNMad}
\begin{widetext}
\begin{equation}
V_{{\rm{ISHE}}} \propto j_\text{s} \left[\cos\theta\cos(\theta-\phi)\int_0^{d_\text{N}} {e^{ - x/{\lambda _{\rm{N}}}}}dx  + \sin\theta\sin(\theta-\phi)\int_0^{d_\text{N}} {\mathop{\rm Re}\nolimits} \left[ {{e^{ - x/{\lambda _\omega }}}} \right] dx\right],\label{hanle}
\end{equation}
\end{widetext}
where $d_\text{N}$ is the thickness of the p-Si layer. From equation (\ref{hanle}), the electromotive force without taking into account spin precession ($\omega_\text{L}\tau_\text{sf} \ll 1$) is given by $V_\text{ISHE}\propto j_\text{s}\cos\phi \int_0^{d_\text{N}} {e^{ - x/{\lambda _{\rm{N}}}}}dx$, which is valid for materials where the spin relaxation time is very fast, such as Pt (Ref. \onlinecite{AndoJAPfull}). Here, calculated $\theta$ dependence of $\Delta V_\text{s}$ is shown in the inset to Fig.~\ref{fig3}f for $\omega_\text{L}\tau_\text{sf} \ll 1$ (the black curve), $\tau_\text{sf}=10$ ps (the blue curve), and $\tau_\text{sf}=20$ ps (the red curve) with $4\pi M_\text{s}=0.852$ T. As $\tau_\text{sf}$ increases, spin precession reduces the electromotive force; the drastic variation of $\Delta V_{\text{s}}$ around $\theta=80^{\circ}$ for $\omega_\text{L}\tau_\text{sf} \ll 1$ becomes gentle for $\tau_\text{sf}=10$ ps and $\tau_ \text{sf}=20$ ps due to spin precession. The experimentally measured $\theta$ dependence of $\Delta V_\text{s}$ is well reproduced using equation~(\ref{hanle}) with $\tau_\text{sf}=9\pm 3$ ps as shown in Fig.~\ref{fig3}f, where $\Delta V_\text{s}$ is obtained from Fig.~\ref{fig3}e. This is the direct evidence of the observation of the ISHE in the p-Si layer; the $\Delta V_\text{s}$ signal cannot be attributed to the ISHE in the Ni$_{81}$Fe$_{19}$ layer, since the spin relaxation time in Ni$_{81}$Fe$_{19}$, $\tau_\text{sf}=9$ fs, is so fast that $\omega_\text{L}\tau_\text{sf}\ll1$ is satisfied (see the dashed curve in Fig.~\ref{fig3}f), where $\tau_\text{sf}$ is obtained from the spin diffusion length~\cite{Bass} $\lambda_\text{F}=3$ nm and diffusion constant~\cite{NiFeDiffusion} $D_\text{F}=10$ cm$^2$s$^{-1}$. This result also supports that magnetogalvanic effects, i.e., the OHE and AHE, and heating effects are irrelevant to $\Delta V_\text{s}$.

\bigskip\noindent
\textbf{Discussion}\\
The above experimental results allow estimation of the spin Hall conductivity of the p-Si layer. In the FMR condition when the magnetic field is applied along the film plane, the magnitude of the ISHE signal $V_\text{ISHE}$ is obtained from equation~(\ref{pumping2}) with the equivalent circuit model of the spin-pumping induced ISHE where short-circuit currents in the Ni$_{81}$Fe$_{19}$ layer are taken into account~\cite{AndoJAPfull}: $V_\text{ISHE}=\left( {2e}/{\hbar} \right)  \left[{w_\text{F} \theta_\text{SHE}\lambda_\text{N}\tanh(d_\text{N}/2\lambda_\text{N})}\right]/\left[{d_\text{N} \sigma_\text{N}+d_\text{F}\sigma_\text{F}}\right]j_\text{s}$. Here, $w_\text{F}$, $d_\text{F}$, and $\sigma_\text{F}$ are the length defined as in Fig.~\ref{fig1}b, thickness, and electric conductivity of the Ni$_{81}$Fe$_{19}$ layer, respectively. Using the parameters for the Ni$_{81}$Fe$_{19}$/p-Si film, $w_\text{F}=2.0$ mm, $D_\text{N}=3.23$ cm$^{2}$s$^{-1}$, $d_\text{N}=4$ $\mu$m, $d_\text{F}=10$ nm, $\sigma_\text{N}=2\times 10^2$ $\Omega^{-1}$cm$^{-1}$, $\sigma_\text{F}=1.5\times 10^4$ $\Omega^{-1}$cm$^{-1}$, $4\pi M_s=0.852$ T, $\alpha= 0.0088$, $h=0.16$ mT, $\tau_\text{sf}=9$ ps, and $\Delta V_\text{s}=0.87$ $\mu$V, we find $g_\text{r}^{\uparrow\downarrow}\theta_\text{SHE}=5.9\times10^{14}$ m$^{-2}$. Here, the spin mixing conductance $g_\text{r}^{\uparrow\downarrow}$ can be obtained from the enhancement of the FMR spectral width due to the spin pumping~\cite{Tserkovnyak_rev}. The spin mixing conductance of the Ni$_{81}$Fe$_{19}$/p-Si film is estimated from the FMR spectral width for the Ni$_{81}$Fe$_{19}$/p-Si film and a Ni$_{81}$Fe$_{19}$/SiO$_{2}$ film as $g_\text{r}^{\uparrow\downarrow}=(4.7\pm0.5) \times 10^{18}$ m$^{-2}$. Thus we obtain the spin Hall angle for the p-Si layer $\theta_\text{SHE}\approx 1\times 10^{-4}$, which corresponds to the spin Hall conductivity $\sigma_\text{SHE}\approx2\times 10^{-2}$ $\Omega^{-1}$cm$^{-1}$. These values are much smaller than those for doped GaAs~\cite{Matsuzaka}, showing that this approach enables high sensitive electric measurement of the ISHE. The successful measurement of the ISHE in silicon is attributed to its high electric resistivity, which is essential for large voltage generation due to the ISHE, and the high density spin injection into macroscopic area by the spin pumping. 

Although spin injection into n-type Si has been reported by several groups~\cite{Jonker,Dash,ando:182105,suzukiAPEX}, there is only one report of room-temperature spin injection into p-type Si, using tunnel contacts~\cite{Dash}. The successful observation of the ISHE in the Ni$_{81}$Fe$_{19}$/p-Si film now confirms this by a different approach, namely, dynamical spin injection. The present experiment shows that the spin relaxation time in the p-Si layer is $\tau_\text{sf}=9$ ps. Here note that, in the direct Ni$_{81}$Fe$_{19}$/p-Si contact, the spin relaxation time near the interface may be reduced because of the coupling of the spins in the p-Si layer to the Ni$_{81}$Fe$_{19}$ layer~\cite{PhysRevB.84.054410}. The spin relaxation time obtained using the electrical spin injection is  $\tau_\text{sf}=270$ ps for p-Si with the doping concentration of $N_\text{A}=4.8\times 10^{18}$ $\text{cm}^{-3}$ (Ref.~\onlinecite{Dash}). Therefore, the spin relaxation time in p-Si obtained by both the electrical and dynamical spin injection is much longer than the momentum relaxation time $\sim 5$ fs in the p-Si layer\cite{10.1063/1.1690487}; understanding the hole spin relaxation in p-type Si remains a challenge.

We revealed that silicon has the potential to be used not only as a spin-current-transmission path~\cite{suzukiAPEX,10.1063/1.3624923} but also as a spin-current detector in spite of its weak spin-orbit interaction. Although the spin/charge current conversion efficiency is not large in the p-Si layer, the spin Hall effects in silicon can now be further explored; the combination of the spin pumping and ISHE paves the only way for quantitative exploration of the spin-orbit coupling effect in silicon for different doping density and dopant type. This approach provides also a way to extract the spin relaxation time $\tau_\text{sf}$; as shown in Fig.~\ref{fig4}d, the magnitude of the electromotive force due to the ISHE is strongly dependent on $\tau_\text{sf}$ under the oblique magnetic field, especially when $\tau_\text{sf}$ is of the order of 10 ps. Furthermore, the approach presented here, thanks to the high-density spin injection, opens the way for exploring the spin Hall effects in a wide range of materials, including high resistivity materials with weak spin-orbit interaction. This extends the range of potential materials for spin-current detector without magnetic materials.

\bigskip\noindent
\textbf{Methods}\\
\textbf{Sample preparation.} The sample used in this study is a Ni$_{81}$Fe$_{19}$/B-doped Si (Ni$_{81}$Fe$_{19}$/p-Si) film with a doping concentration of $N_\text{A}=2\times 10^{19}$ $\text{cm}^{-3}$. Two 30-nm-thick AuPd electrodes were sputtered on a silicon-on-insulator (SOI) substrate (see Fig.~\ref{fig1}a) in an Ar atmosphere. After the sputtering, the SOI substrate was annealed at 400 $^\circ$C for 10 minutes in a high vacuum, which yields ohmic contacts to the p-Si layer (see Fig.~\ref{fig1}b). The 10-nm-thick Ni$_{81}$Fe$_{19}$ layer was then deposited on the p-Si layer by electron-beam evaporation in a high vacuum. Immediately before the evaporation, the surface of the p-Si layer was cleaned by Ar ion etching. The surface of the Ni$_{81}$Fe$_{19}$ layer and AuPd contact is of a $1.0\times 2.0$ mm$^2$ rectangular shape and of a $1 \times 0.5$ mm$^2$ rectangular shape, respectively. The distance from the AuPd contact to the Ni$_{81}$Fe$_{19}$ layer is $\sim 300$ $\mu$m.

\bigskip\noindent
\textbf{Electric voltage due to inverse spin Hall effect.} The observed electromotive force in the Ni$_{81}$Fe$_{19}$/p-Si film is the combination of the inverse spin Hall effect (ISHE) in the p-Si layer, the ordinary Hall effect (OHE) in the p-Si layer, the anomalous Hall effect (AHE) in the Ni$_{81}$Fe$_{19}$ layer, and heating effects. The direct contribution from the ISHE in the p-Si layer can be extracted as follows. The OHE in the p-Si layer induces a d.c. electromotive force from an a.c. charge current due to a microwave electric field and an a.c. stray field due to precessing magnetization. Notable is that this rectified voltage changes its sign across the resonance field, i.e., the sign of the electromotive force when $H<H_\text{FMR}$ is opposite to that when $H>H_\text{FMR}$, since the phase of magnetization precession shifts by $\pi$ at resonance. Therefore, the shape of the electromotive force due to the OHE is asymmetric as shown in Fig.~\ref{fig2}g. Here, a microwave magnetic field cannot create a detectable d.c. OHE voltage, since the direction of the microwave magnetic field is parallel to the direction across the electrodes (the $y$ direction, see Fig.~\ref{fig1}a). Furthermore, the microwave magnetic field is independent on the FMR. The shape of the electromotive force due to the AHE is also asymmetric, since it is a rectified voltage induced by the combination of a charge current due to a microwave electric field and precessing magnetization~\cite{AndoJAPfull}. In contrast to the rectified voltage due to the OHE and AHE, the electromotive force due to the ISHE is proportional to the intensity of microwave absorption~\cite{ando:262505}. This indicates that the shape of the electromotive force due to the ISHE is symmetric as shown in Fig.~\ref{fig2}g and thus the electromotive force due to the OHE and AHE can be eliminated from the observed electromotive force. The electromotive force due to the sample heating is induced by the Seebeck effect, which is independent on the magnetic field direction. The Seebeck effect in the Ni$_{81}$Fe$_{19}$/p-Si film is induced by a lateral temperature gradient along the film plane due to small but finite misalignment of the position of the Ni$_{81}$Fe$_{19}$ layer with respect to the substrate. In fact, the magnitude of the symmetric component of $V$ which does not change the sign with reversal of ${\bf H}$ depends strongly on samples. A longitudinal temperature gradient, i.e., a temperature gradient perpendicular to the film plane, may induce a voltage via the Nernst effect. Although the Nernst effect induces a ${\bf H}$ dependent voltage, Fig.~\ref{fig3}f clearly shows that this effect is irrelevant to the $\Delta V_\text{s}$ signals; the variation of $\Delta V_\text{s}$ cannot be reproduced by the cross product of a longitudinal temperature gradient and the external magnetic field. The $\theta$ dependence of $\Delta V_\text{s}$ is well reproduced using equation~(\ref{hanle}) with the spin relaxation time $\tau_\text{sf}=9$ ps for $P_\text{MW}=100$, 150, and 200 mW. All errors and error bars represent the 90\% confidence interval.

\bigskip\noindent
\textbf{Equivalent circuit model.} As shown in Fig.~\ref{fig4}a, $j_{\text{s},x}^z (x)$ decays because of the spin relaxation in the p-Si layer; the charge current density $j_\text{c}(x)$ generated by the ISHE also depends on $x$, which induces short circuit currents in the Ni$_{81}$Fe$_{19}$ and p-Si layers~\cite{AndoJAPfull}. Here, a total charge current generated by the ISHE is $I_\text{N}=\int ^{d_\text{N}}_0 j_\text{c}(x)dx$, where $d_\text{N}$ is the thickness of the p-Si layer. By dividing the p-Si layer into $n$ layers, an equivalent circuit of the Ni$_{81}$Fe$_{19}$/p-Si film is obtained as shown in Fig.~\ref{fig4}b, where $R_\text{F}$ is the electrical resistance of the Ni$_{81}$Fe$_{19}$ layer. The electrical resistance $R_\text{N}^i$ and the charge current $I_\text{N}^i$ generated by the ISHE of the $i$th layer satisfy $R_\text{N}^{-1}=\sum^n_{i=1} (R_\text{N}^i)^{-1}$ and $I_\text{N}=\sum^n_{i=1} I_\text{N}^i$, where $R_\text{N}$ is the electrical resistance of the p-Si layer. It is straightforward to convert the circuit shown in Fig.~\ref{fig4}b into that shown in Fig.~\ref{fig4}c. Thus, the electromotive force due to the ISHE is obtained as $V_\text{ISHE}=[R_\text{F}R_\text{N}/(R_\text{F}+R_\text{N})]I_\text{N}$. This result shows that $V_\text{ISHE}$ is proportional to the total charge current $I_\text{N}$ generated by the ISHE; the electromotive force due to the ISHE in the Ni$_{81}$Fe$_{19}$ layer is negligibly small because of the extremely short spin-diffusion length~\cite{Bass} and small spin-Hall angle~\cite{NiFeAHE}.

\bigskip\noindent
\textbf{Acknowledgements}\\
The authors thank to S. Takahashi and R. Takahashi for valuable discussions. This work was supported by the Cabinet Office, Government of Japan through its ``Funding Program for Next Generation World-Leading Researchers," the Asahi Glass Foundation, the Casio Foundation, the Kurata Foundation, and JST-CREST ``Creation of Nanosystems with Novel Functions through Process Integration".

\bigskip\noindent
\textbf{Author contributions}\\
K.A. designed the experiment, collected all of the data, performed analysis of the data, and wrote the manuscript. E.S. supervised the study. Both authors discussed the results and commented on the manuscript.

\bigskip\noindent
\textbf{Additional information}\\
The authors declare no competing financial interests.

\end{document}